# New algorithm for astrometric reduction of the wide-field images

Volodymyr Akhmetov
*Laboratory of Astrometry*
*Institute of Astronomy,*
*V. N. Karazin Kharkiv National University*
Kharkiv, Ukraine
akhmetovvs@gmail.com

Sergii Khlamov
*Laboratory of Astrometry*
*Institute of Astronomy,*
*V. N. Karazin Kharkiv National University*
Kharkiv, Ukraine
sergii.khlamov@gmail.com

Vladislav Khramtsov, Artem Dmytrenko
*Department of Astronomy and Space Informatics,*
*V. N. Karazin Kharkiv National University*
Kharkiv, Ukraine
astronom.karazin007@gmail.com

*Abstract* — In this paper we presented the modified algorithm for astrometric reduction of the wide-field images. This algorithm is based on the iterative using of the method of ordinary least squares (OLS) and statistical Student t-criterion. The proposed algorithm provides the automatic selection of the most probabilistic reduction model. This approach allows eliminating almost all systematic errors that are caused by imperfections in the optical system of modern large telescopes.

*Keywords* — database, big data, catalogue, astrometry, reduction, wide field.

## I. Introduction

In recent years, with the growth of technological capabilities and computer technology, the automation and computerization of the processing has become more widespread in the different areas of the human activity. Astronomy, in particular astrometry, is not an exception too. Because of the construction of large ground-based telescopes, launching of the space missions, there is a very rapid increasing of the amount of observational data. These data at the first stage are represented as images of the stellar sky that should be processed, measured and cataloged.

For this purpose, first of all, it is necessary to perform an astrometric reduction. The last one should be performed as accurately as possible because it can be used in the different software for automated image processing, such as CoLiTec (Collection Light Technology) software (http://neoastrosoft.com) [1, 2], Astrometrica [3] and others. That is why the highest accuracy level of the astrometric reduction is necessary in the modern astronomy.

In the most cases, for performing the astrometric reduction the classical computational methods are used, in particular the method of ordinary least squares (OLS) [4].

As we know, for a sufficiently large number of iterations, as in our case of astrometric reduction, the estimates by the OLS method coincide with the estimates by maximum likelihood method [5]. However, the last one has some disadvantages:

- incomplete evidence for using the maximum likelihood criterion when some parameters are unknown [4, 6];
- necessary to select the value of boundary decisive statistics [6];

Therefore, the authors suggest also using the statistical criterions, such as Student *t*-criterion or Fisher *f*-criterion, instead of using the maximum likelihood criterion, as in the works [7, 8].

Until recently, when the telescopes were not so advanced, the field of view (FOV) of telescopes was not so wide and the number of the reference stars was measured by the first dozens.

Presently, when the modern telescopes cover large areas of the sky and have wide FOV, there are from tens to several hundred thousands of the reference objects will be on the image. This allows performing the astrometric reduction with the highest accuracy level and eliminating all systematic errors caused by imperfections in the optical system. Therefore, it is necessary to improve computational methods that will make possible to perform the assigned researching tasks with a high level of accuracy during the minimum of computational time.

In this paper we propose the new algorithm for astrometric reduction of the wide-field images into the system of the modern astronomical catalogues. Proposed algorithm is based on the iterative using of the OLS method and statistical Student *t*-criterion, which allows with the high accuracy to determine all significant coefficients of decomposition that best describe the systematic errors contained in the measured positions of objects.

## II. Astrometric reduction

The classic astrometric reduction was developed for the classic type astrographs with a relatively small and flat FOV (see, for example, König A. [9]). These telescopes provide the image of the stellar sky, which with a good approximation can be regarded as the central projection of the sphere onto the plane. For such telescopes, the basic concepts of Gaussian optics are fair.

However, for super wide-angle and super high-aperture systems such as the modern large telescopes Thirty Meter Telescope (TMT) [10] and the Large Synoptic Survey Telescope (LSST) [11], the concepts of geometrical optics lose their meaning. Nevertheless, in this case the central projection is the closest mathematical model of real astrophotography too. With help of this model, we still can solve many astrometric tasks.

In the each specific case, it is necessary to choose the appropriate reduction model, and to complement it with the necessary terms for minimization of the discrepancies between the model and the real image of the stellar sky.

So, the systems of conditional equations for determination of the parameters of reduction model can be presented in the array form [12]:

$$X_{Nm} A_{m1} = \Xi_{N1}, \quad (1)$$

$$X_{Nm} E_{m1} = \Theta_{N1}, \quad (2)$$

where the $X_{Nm}$ array has $N$ rows and $m$ columns that are predetermined by the coefficients (measured coordinates of N reference stars) of the conditional equations; $A_{m1}$, $E_{m1}$ – the vectors that are composed of the predetermined parameters $a_i$ and $e_i$; $i = 1, 2,..., m$; $\Xi$ and $\Theta$ – the vectors that are composed of the tangential coordinates $\xi_i$, $\eta_i$ of reference stars.

Solving the system of conditional equations using the OLS method provides the most probable unbiased values of unknown parameters:

$$\tilde{A}_{m1} = C^{-1}_{mm} X_{mN} \Xi_{N1}, \quad (3)$$

$$\tilde{E}_{m1} = C^{-1}_{mm} X_{mN} \Theta_{N1}, \quad (4)$$

where $\tilde{A}_{m1}$ and $\tilde{E}_{m1}$ – vectors of the estimates of unknown parameters; $a_i$ and $e_i$; $i = 1, 2,..., m$; $C^{-1}_{mm}$ – symmetric array that is inverse to the matrix of normal equations:

$$C_{mm} = X_{mN} X_{Nm}. \quad (5)$$

Objects estimates $\tilde{A}_{m1}$ and $\tilde{E}_{m1}$ are determined taking into account the weighting of the tangential coordinates that depend of the accuracy of their spherical coordinates.

The estimates of reduction errors that are based on the internal convergence of the initial data with taking into account the correlations of the parameters $a_i$ and $e_i$ is represented as:

$$\sigma = \sqrt{\sum v_i^2 / N - m}. \quad (6)$$

The correlation arrays of the parameters $a_i$ and $e_i$:

$$K(a_i\, a_i), \quad (7)$$

$$K(e_i\, e_i). \quad (8)$$

The influence of the configuration of reference and determined stars are estimated by the following equation:

$$X^0_{1m} C^{-1}_{mm} X^0_{m1}, \quad (9)$$

where $X^0_{1m} = \| 1, x_0, y_0, x_0^2, ... x_0^{pq}, y_0^{rs} \|$; $X^0_{m1} = (X^0_{1m})^T$.

Parameter $X^0_{1m}$ allows noticing and correcting of the individual contribution of each reference star into the total reduction error. The index $^0$ indicates the measured coordinates of the investigated object.

By using the Student $t$-criterion, we can estimate the significance of the expansion for truncated model:

$$t = (x' - \mu) \sqrt{m / s}, \quad (10)$$

where $x'$ is the selective mean and equals to:

$$x' = 1/m \sum x_i \quad (11)$$

and $s^2$ is the selective dispersion and equals to:

$$s^2 = 1/(m - 1) \sum (x_i - x')^2, \quad (12)$$

When using the classical OLS method for the performing a reduction, it is necessary to know in advance degree and type of the reduction model. This model should best describe all distortions caused by the optical system of the telescope. In advance, it is not always possible to choose the degree of a polynomial that will describe all the systematic components.

Also, when determining the decomposition coefficients using only one iteration, all random errors that hide in the data due to imperfections in the optical system, will be not discarded from the used data. This will lead to a distortion of the final result and incorrect interpretation of the obtained data.

III. NEW ALGORITHM OF THE COORDINATE REDUCTION

We offer an improved algorithm of the OLS method, one of the features of which is the automatic selection of the degree and shape of the reduction polynomial, which will best describe the experimental data.

There are two equations for each reference object. Because of the large number of the reference objects $N$, which sometimes reaches hundreds of thousands, the solving a system of equations with high degree polynomials directly through the $X_{mN}$ array and its transposed $X_{Nm}$ array is not justified.

Therefore, we immediately create an array $C_{mm}$ of the normal equations and a vector with multiplying of the tangential coordinates and $X_{mN}$:

$$\begin{pmatrix} n & \sum_{i=1}^{n} x_{i1} & \cdots & \sum_{i=1}^{n} x_{ip} \\ \sum_{i=1}^{n} x_{i1} & \sum_{i=1}^{n} x_{i1}^2 & \cdots & \sum_{i=1}^{n} (x_{i1} \cdot x_{ip}) \\ \cdots & \cdots & \cdots & \cdots \\ \sum_{i=1}^{n} x_{ip} & \sum_{i=1}^{n} (x_{i1} \cdot x_{ip}) & \cdots & \sum_{i=1}^{n} x_{ip}^2 \end{pmatrix} \begin{pmatrix} \sum_{i=1}^{n} y_i \\ \sum_{i=1}^{n} (y_i \cdot x_{i1}) \\ \cdots \\ \sum_{i=1}^{n} (y_i \cdot x_{ip}) \end{pmatrix}$$

Such an approach makes it possible to quickly obtain estimates of all coefficients of the polynomial reduction by OLS method for a large number of the reference objects. At the first iteration of the solution, we use a polynomial of degree 5, which provides 21 constant plates. Also, we get a reduction error, which we use in the next iteration to eliminate reference objects whose residuals more then $3\sigma$.

After several iterations of getting rid of the noisy objects, in our OLS method we apply the statistical Student $t$-criterion of reliability. This allows getting rid of the some insignificant coefficients of the reduction model, thereby increasing the accuracy of determining the significant one. The elimination of insignificant members of a polynomial is also carried out several times, until the general residual $\sigma$ will not reach a minimum. As practice shows, in most cases, for this you need to perform 7-12 iterations.

The main advantage of the proposed approach is that we can obtain high-precision values of the desired coefficients of expansion. They will be much more accurate than if they were obtained using a classical approach, which uses only one iteration with the reduction model that is not always correct.

The distribution of residuals after performing the astrometric reduction using the standard reduction model (third-degree polynomial) for Schmidt telescopes is shown in figure 1.

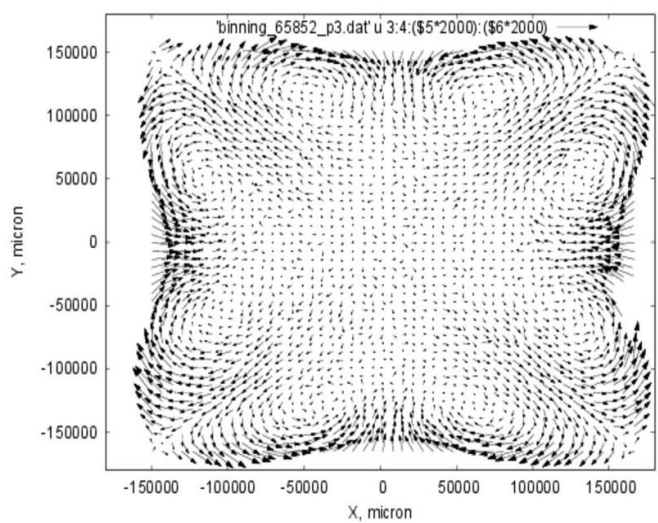

Fig. 1. Mean vector residual map using a standard third-degree-order polynomial plate model.

Unfortunately, we can see a significant systematics residual as a function of stellar position on frame. This indicates to the incorrect selection of the reduction model and, accordingly, the incomplete elimination of systematic errors in the measured positions of the reference objects.

Figure 2 shows the distributions of residuals after using our method by the fifth-degree polynomial. Any significant symmetric errors in the measured positions of the reference objects are absent after performing the astrometric reduction using our method by the fifth-degree polynomial.

This approach allows us to performing the astrometric reduction for wide-field images with the highest accuracy level and excludes all coordinate-dependences errors in the observational data.

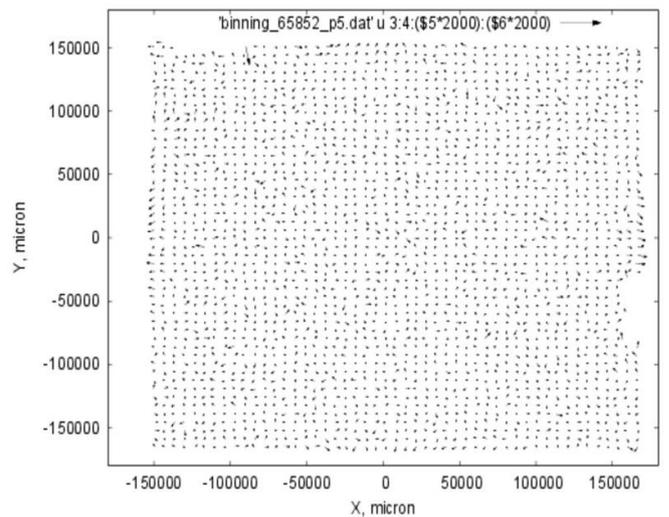

Fig. 2. Mean vector residual map using our method by the fifth-degree-order polynomial plate model.

## IV. REDUCTION OF SUPERCOSMOS DATA

We created the software for performing the astrometric reduction of the big sets of data. The software is based on C++ realization. Using the developed software the reduction of all measured positions of objects on the digitized photographic plates of SuperCOSMOS survey [13] into the reference PMA catalogue [14] was performed. The PMA catalogue contains about 421 million of the reference objects up to 21 G magnitude [15].

There are from tens to several hundred thousands of the reference objects on each digitized photographic plate of the SuperCOSMOS survey [16]. FOV of each such plate is about 6 x 6 degrees, which is commensurable with the working FOV of the LSST telescope.

There are more than 5978 such digitized photographic plates of the SuperCOSMOS survey in different photometric filters (B, R, I). All observations from these plates were performed in the different observational epochs for both the northern and southern hemispheres and provided in the table 1.

TABLE I.  SUPERCOSMOS SURVEYS THAT ARE USED IN THIS WORK TO DERIVING POSITIONS OF THE OBJECTS AT AN EPOCH OF THEIR OBSERVATION

| Survey name | SERC-J | SERC-R | SERC-I | POSS I-R | POSS II-B | POSS II-R | POSS II-I |
|---|---|---|---|---|---|---|---|
| Epoch min | 1974.463 | 1984.546 | 1978.890 | 1949.112 | 1985.000 | 1986.000 | 1987.000 |
| Epoch max | 1994.856 | 2001.289 | 2003.000 | 1957.970 | 2002.208 | 1999.000 | 2002.000 |
| FOV, ° | 6 x 6 | 6 x 6 | 6 x 6 | 6 x 6 | 6 x 6 | 6 x 6 | 6 x 6 |
| DEC min, ° | -90 | -90 | -90 | 2 | 2 | 2 | 2 |
| DEC max, ° | 3 | 3 | 3 | 90 | 90 | 90 | 90 |
| Fields | 894 | 894 | 894 | 824 | 824 | 824 | 824 |
| Color | Blue | Read | Near-IR | Red | Blue | Red | Near-IR |
| Mag limit | 23.0 | 22.0 | 19.0 | 20.0 | 22.5 | 20.8 | 19.5 |

As it turned out, the use of a large number of high-precision positions of the reference objects it possible to detect the coordinate-photometric dependence of some of the reduction model coefficients. This dependence is called a regular magnitude equation. Figure 3 shows the values of this coefficient of the reduction model as a function of the stellar magnitude. The error of the obtained coefficients of the reduction model depends on the stellar magnitude. Only in the brightest and weakest region where the number of reference objects is not enough, we have received large errors in determining the reduction coefficients.

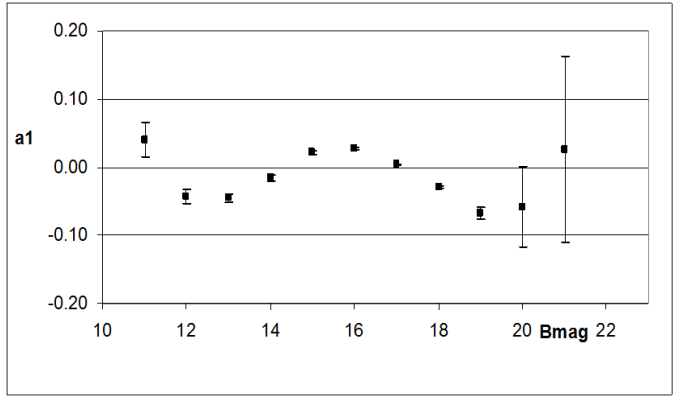

Fig. 3. Significant coefficient $a1$ as a function of stellar magnitude radius.

Using proposed method for astrometric reduction, the high-accurate positions for all objects of SuperCOSMOS survey were calculated. Total count of these objects for both the northern and southern hemispheres is more than 1 billion. These data and proposed method are very helpful in the process of creating the high-density astrometric and photometric catalogue in the modern big data surveys. For created a high-density reference frame the new reduction data of the SuperCOSMOS was used [17] as well as an especial tool for the cross-match [18].

## V. CONCLUSIONS

In this paper, we developed the new modified algorithm for astrometric reduction of the wide-field images using the iterative method of the ordinary least squares (OLS). We created software to perform astrometric reduction in the different ranges of stellar magnitudes. This software allows excluding of all coordinate-photometric errors in the measured positions of objects even at the edges of wide-field images.

Using of the developed method has shown that when performing the reduction of large astronomical catalogues, the new proposed algorithm allows performing the reduction into the system of reference catalogue with the highest accuracy level. Also, the proposed algorithm suppresses all random noise that is presented in the input data and is caused by imperfections in the optical system of modern large telescopes.

ACKNOWLEDGMENT

This research has made use of data obtained from the SuperCOSMOS Science Archive, prepared and hosted by the Wide Field Astronomy Unit, Institute for Astronomy, University of Edinburgh, which is funded by the UK Science and Technology Facilities Council.